\documentclass[aps,prb,10pt,twocolumn,superscriptaddress]{revtex4-2}

\usepackage[english]{babel}

\usepackage[ascii]{inputenc}
\usepackage{amsmath}
\usepackage{graphicx}
\usepackage{physics}
\usepackage{braket}
\usepackage{tikz}
\usetikzlibrary{fit}
\usetikzlibrary{positioning}
\usetikzlibrary{decorations.pathmorphing,decorations.pathreplacing}
\usepackage{tikz-network}
\usepackage[caption=false]{subfig}
\usepackage[colorlinks=true, allcolors=blue]{hyperref}
\usepackage{makecell}
\usepackage{afterpage}

\graphicspath{{pictures/}}
\makeatletter
\def\input@path{{tikz_figures/}}
\makeatother

\pgfdeclaredecoration{simple line}{start}
{
  \state{start}[width = +0pt,
                next state=step]{
    \pgfpathmoveto{\pgfpoint{0pt}{0pt}}
  }
  \state{step}[auto end on length    = 3pt,
               auto corner on length = 3pt,               
               width=+1pt]
  {
    \pgfpathlineto{\pgfpoint{1pt}{0pt}}
  }
  \state{final}
  {}
}

\makeatletter
\def\pgf@stroke@inner@line@if@needed{%
  \ifdim\pgfinnerlinewidth>0pt\relax%
    \let\pgf@temp@save=\pgf@strokecolor@global
    \pgfsys@beginscope%
    {%
      \pgfsetrectcap
      \pgfsys@setlinewidth{\pgfinnerlinewidth}%
      \pgfsetstrokecolor{\pgfinnerstrokecolor}%
      \pgfsyssoftpath@invokecurrentpath%
      \pgfsys@stroke%
    }%
    \pgfsys@endscope%
    \global\let\pgf@strokecolor@global=\pgf@temp@save
  \fi%
}

\definecolor{anc_color}{rgb}{0.8941176470588236, 0.10196078431372549, 0.10980392156862745}
\definecolor{phys_color}{rgb}{0.17254901960784313, 0.6274509803921569, 0.17254901960784313}
\definecolor{contr_color}{rgb}{0.12156862745098039, 0.4666666666666667, 0.7058823529411765}
\definecolor{fill_left}{rgb}{0.7764705882352941, 0.8588235294117647, 0.9372549019607843}
\definecolor{fill_center}{rgb}{0.9921568627450981, 0.8156862745098039, 0.6352941176470588}
\definecolor{fill_right}{rgb}{0.7764705882352941, 0.8588235294117647, 0.9372549019607843}
\definecolor{left_rect}{rgb}{0.12156862745098039, 0.4666666666666667, 0.7058823529411765}
\definecolor{right_rect}{rgb}{0.12156862745098039, 0.4666666666666667, 0.7058823529411765}
\definecolor{extra_color}{rgb}{1.0,0.8509803921568627,0.1843137254901961}

\begin{document}

\title{Block encoding of matrix product operators}
\author{Martina Nibbi}
\email{martina.nibbi@tum.de}
\affiliation{Technical University of Munich, School of CIT, Department of Computer Science, Boltzmannstra{\ss}e 3, 85748 Garching, Germany}

\author{Christian B.~Mendl}
\email{christian.mendl@tum.de}
\affiliation{Technical University of Munich, School of CIT, Department of Computer Science, Boltzmannstra{\ss}e 3, 85748 Garching, Germany}
\affiliation{Technical University of Munich, Institute for Advanced Study, Lichtenbergstra{\ss}e 2a, 85748 Garching, Germany}

\begin{abstract}
Quantum signal processing combined with quantum eigenvalue transformation has recently emerged as a unifying framework for several quantum algorithms.
In its standard form, it consists of two separate routines: block encoding, which encodes a Hamiltonian in a larger unitary, and signal processing, which achieves an almost arbitrary polynomial transformation of such a Hamiltonian using rotation gates.
The bottleneck of the entire operation is typically constituted by block encoding and, in recent years, several problem-specific techniques have been introduced to overcome this problem.
Within this framework, we present a procedure to block-encode a Hamiltonian based on its matrix product operator (MPO) representation.
More specifically, we encode every MPO tensor in a larger unitary of dimension $D+2$, where $D = \lceil\log(\chi)\rceil$ is the number of subsequently contracted qubits that scales logarithmically with the virtual bond dimension $\chi$. 
Given any system of size $L$, our method requires $L+D$ ancillary qubits in total, while the number of one- and two-qubit gates decomposing the block encoding circuit scales as $\mathcal{O}(L\cdot\chi^2)$.
Moreover, block encodings generally require post-selection. Our model exhibits a success probability decaying exponentially with $L$. Nevertheless, we present a way to mitigate this limitation for finite systems.
\end{abstract}

\maketitle

\section{Introduction}
Quantum signal processing (QSP) combined with ``qubitization'' provides a unifying paradigm for several quantum algorithms \cite{Gilyen2019, Low_2019, Martyn_2021, Kikuchi_2023}. The quantum eigenvalue transformation (QET) technique is part of this framework. It facilitates a tailored polynomial transformation of the eigenvalues of a Hermitian matrix, which is typically a Hamiltonian $\mathcal{H}$ of a quantum system of interest. QET has versatile applications, like eigenspace filtering, ground state preparation, and Hamiltonian simulation.

In this process, the most significant bottleneck is represented by the block encoding of $\mathcal{H}$. It is well known that a unitary operator embedding $\mathcal{H}$ exists. However, decomposing such a unitary efficiently into one- and two-qubit gates requires some extra knowledge of the system and problem-specific techniques.
In this context, we propose a versatile block encoding method based on a matrix product operator (MPO) representation of $\mathcal{H}$, where such an MPO form is typically known or can be constructed for systems of interest. Our method uses unitary dilations of the individual MPO tensors. 
Our approach achieves a block encoding with a number of one- and two-qubit gates scaling only linearly with the system size $L$ and quadratically with the virtual bond dimension $\chi$ of the MPO, while the number of ancillary qubits scales like the sum of $L$ and $\log(\chi)$. 

The post-selection cost must also be accounted for in order to get a full picture of the computational cost. All block encoding methods share this issue, and it poses a challenge to the applicability of our algorithm, in particular, due to exponential decay in the system size. 
We will discuss this aspect in detail throughout the text and provide some ideas on how to overcome this limitation for finite systems.

Our theoretical justification starts then in section \ref{sec_TN}, where we provide a concise definition of MPOs within the context of tensor networks. This quick introduction is followed by the description of our algorithm in section \ref{sec_circuit_rep}, which encodes such MPOs on a quantum computer.
This task requires the use of ancillaries,  measurements and post-selection, as the MPO tensors are non-unitary in general.

We discuss the application of our method for QET in section \ref{sec_QET} and, specifically, we outline two methods to implement the signal processing operator: the original one described by Martyn et al.~\cite{Martyn_2021} and an alternative that requires no auxiliary qubits.
The theoretical proof of the correctness of this method is provided in appendix~\ref{appendix_signal_proc}.

We then perform a cost analysis in terms of one- and two-qubit gates count, comparing our method to the linear combinations of unitaries (LCU) block encoding technique in section \ref{sec_cost_analysis}. 
The results show that our technique is characterized by a scaling of $\mathcal{O}(L\cdot\chi^2)$,  while the cost of LCU varies depending on the number of unitaries composing the Hamiltonian. We also compare, on a theoretical level, the post-selection cost for both MPO and LCU block encodings, while a numerical analysis for systems of interest can be found in appendix~\ref{appendix_p_success}.

Finally, we present some practical applications in section \ref{sec_applications}. We examine the MPO representation of the Ising and Heisenberg Hamiltonians and an exponentially decaying XY model, which exhibits a significant advantage in gate decomposition cost compared to the use of LCU.
We also consider a spinless one-dimensional Fermi-Hubbard model as an example of a fermionic system.
At last, we construct a general tensor product of sums of Paulis as a straightforward use case for realizing the entire QET circuit simulating an eigenstate filtering process.

\section{Tensor networks and matrix product operators}\label{sec_TN}

We consider a physical system that can be represented as a one-dimensional chain with $L$ sites.
Then, a matrix product operator (MPO) is any operator acting on such a system as \cite{Schollwoeck_2011, Zaletel_2015}:
\begin{equation}\label{eq_mpo_def}
    O = \tilde{A}^{(1)}A^{(2)}A^{(3)}\cdots A^{(L-1)}\tilde{A}^{(L)}
\end{equation}
where every term $A^{(\ell)}$ is a matrix of operators acting on the $\ell$-th site of the lattice except for the first and last term, which are respectively a row and column vector of operators.
For concreteness, we will assume that the MPO has uniform virtual bond dimension $\chi$ and physical wires of dimension 2. Other scenarios follow by a straightforward generalization.

Figure~\ref{fig:MPO_general} represents the tensor diagram of Eq.~\eqref{eq_mpo_def}. In Fig.~\ref{fig:MPO_boundary}, we defined the boundary tensors $\tilde{A}^{(1)}$ and $\tilde{A}^{(L)}$ as the contraction between two tensors $A^{(1)}$ and $A^{(L)}$ of degree four, respectively with a row and column vector $R$ and $C$. 
Such a manipulation of the boundary tensors conveniently allows us to work with an MPO with uniform dimensions, while the role of the vectors $R$ and $C$ will be further investigated in sections \ref{sec_QET} and \ref{sec_applications}.

Such a decomposition can always be found through iterated singular values decompositions, exactly like matrix product states (MPS) \cite{Schollwoeck_2011}.
However, this method quickly becomes computationally too expensive for increasing $L$, so it is considered impracticable for most applications.
Instead, it is possible to build the MPO by representing the Hamiltonian through finite state automata \cite{Fr_wis_2010, Crosswhite_2008}. This method is particularly suitable for translation-invariant systems. We will show a few examples in section \ref{sec_applications}.
From now on, we will assume that the MPO decomposition of the target Hamiltonian is known.

\section{Circuit representation}\label{sec_circuit_rep}
Now that the basic concepts for understanding matrix product operators have been introduced, we describe how to implement them on a quantum circuit.
First of all, to make the tensors $A^{(\ell)}$ act on qubits as a quantum gate, we need to reshape them from four-tensors to matrices as shown below: 
\begin{center}
    \includegraphics[width=0.6\columnwidth]{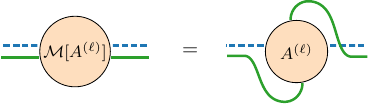}
\end{center}
where we define $\mathcal{M}[A^{(\ell)}]$ as the matrix reshape of the tensor $A^{(\ell)}$.

Given then the virtual bond dimension of the original MPO $\chi$, which we assumed to be uniform, we proceed to expand the matrices $\mathcal{M}[A^{(\ell)}]$ so that their dimensions reach the next power of 2. 
Examples of this procedure will be shown in the applications section.
Then, since we have postulated the physical wire to have dimension 2, we can assume $\mathcal{M}[A^{(\ell)}]$ to have dimensions $2^{D+1}\times2^{D+1}$, where $D$ is defined as:
\begin{equation}\label{eq_virtual_bond_dimension}
    D = \lceil \log(\chi) \rceil
\end{equation}
where the logarithm is intended in base $2$.
It should be clear that if the physical wire can be represented with one single qubit, the virtual bond needs $D$ qubits. From now on, we will refer to the latter as \emph{virtual bond ancillaries}.

\begin{figure}
\centering
\subfloat[]{%
    \includegraphics[width=0.8\columnwidth]{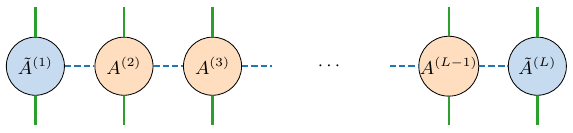}
    \label{fig:MPO_general}}\\
\subfloat[]{%
    \includegraphics[width=\columnwidth]{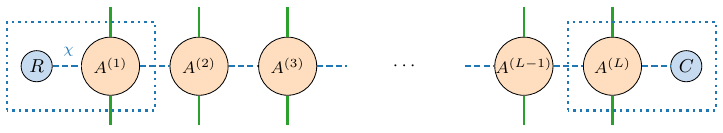}
    \label{fig:MPO_boundary}}\\
\subfloat[]{%
    \includegraphics[width=\columnwidth]{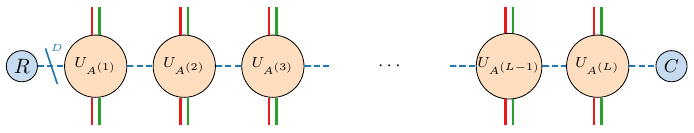}
    \label{fig:MPO_block_encoding}}
\caption{\ref{fig:MPO_general}: Tensor diagram of a general matrix product operator with open boundary conditions. \ref{fig:MPO_boundary}: The three-legged tensors  $\tilde{A}^{(1)}$ and $\tilde{A}^{(L)}$ result from contracting the four-legged tensors $A^{(1)}$ and $A^{(L)}$ respectively with the row and column vectors $R$ and $C$. \ref{fig:MPO_block_encoding}: Unitary dilation of the $A^{(\ell)}$ tensors. The $D$ virtual bond ancillaries are drawn in blue and dashed, the dilation ancillaries in red, and the physical qubits in green.}
\label{fig:MPO}
\end{figure}

The last step needed to represent $\mathcal{M}[A^{(\ell)}]$ as a quantum circuit is a unitary dilation, i.e., we embed each $\mathcal{M}[A^{(\ell)}]$ as a block in a larger unitary matrix. This procedure will require another ancillary qubit, which we will refer to as \emph{dilation ancillary}.
From a diagrammatic point of view, after reshaping the unitary matrices into tensors of degree 4, we get the final MPO form as in Fig.~\ref{fig:MPO_block_encoding}. Here, each green wire corresponds to one physical qubit, each red wire has the role of dilation ancillary, and the $D$ blue dashed wires represent the virtual bond ancillaries.

Using one ancillary qubit has the effect of doubling the size of the encoding unitaries: more specifically, we want to define a unitary matrix $U_{A^{(\ell)}}$ so that its upper-left block is equal to $\mathcal{M}[A^{(\ell)}]$ up to a normalization factor $N_\ell$:
\begin{equation}
\label{eq_block_encoding}
U_{A^{(\ell)}} = \begin{pmatrix}
\mathcal{M}[A^{(\ell)}]/N_\ell &  C\\
B & D
\end{pmatrix}.
\end{equation}
We denote such unitaries as \emph{block encodings} of the matrices $\mathcal{M}[A^{(\ell)}]$.
Note that the normalization factor is necessary because $\mathcal{M}[A^{(\ell)}]$ is typically not normalized and is characterized by the relation:
\begin{equation}\label{eq_normalization_def}
    N_\ell \geq \norm{\mathcal{M}[A^{(\ell)}]},
\end{equation}
with $\norm{\cdot}$ denoting the spectral norm.
Our aim now is to find the blocks $B$, $C$ and $D$ so that, given $A^{(\ell)}$, $U_{A^{(\ell)}}$ is unitary.
We rely on a similar procedure as proposed in \cite[Appendix~D]{Lin_2021}.
\begin{enumerate}
    \item From the unitary constraint on $U_{A^{(\ell)}}$ it follows that 
    \begin{equation} \label{eq_Cholesky}
        B^\dagger B = I -\mathcal{M}[A^{(\ell)}]^\dagger \mathcal{M}[A^{(\ell)}] / N_\ell^2.
    \end{equation}
    Let us consider then the singular value decomposition of the matrix $\mathcal{M}[A^{(\ell)}]$:
    \begin{equation}\label{eq_svd}
        \mathcal{M}[A^{(\ell)}] = U\Sigma V^\dagger
    \end{equation}
    with $U$ and $V$ unitary matrices and $\Sigma$ the diagonal matrix of singular values.
    
    By substituting Eq.~\eqref{eq_svd} into Eq.~\eqref{eq_Cholesky} we get then:
    \begin{equation}
        B^{\dagger}B = V \left(I-\Sigma^2/N_\ell^2\right) V^\dagger.
    \end{equation}
    The normalization factor ensures that the matrix $I-\Sigma^2/N_{\ell}^2$ is non-negative, so that we can identify:
    \begin{equation}
        B = \sqrt{I-\Sigma^2/N_\ell^2} \;V^{\dagger}.
    \end{equation}

    \item Let us consider now the following rectangular matrix:
    \begin{equation}
        W = \begin{pmatrix}
        \mathcal{M}[A^{(\ell)}]/N_\ell \\
        B 
    \end{pmatrix}
    \end{equation}
    where $B$ has been found through the previous step.
    A full QR decomposition applied to $W$ returns a unitary matrix $U$ and a rectangular, upper triangular matrix $R$ so that
    \begin{equation}
        W = U R.
    \end{equation}
    In the first half of its columns, $U$ will share the same entries as $W$, since they were already orthonormal; as a result, the unitary matrix $U$ is of the form~\eqref{eq_block_encoding}, so that we can set $U_{A^{(\ell)}} = U$.
\end{enumerate}

\begin{figure}
    \centering
    \includegraphics{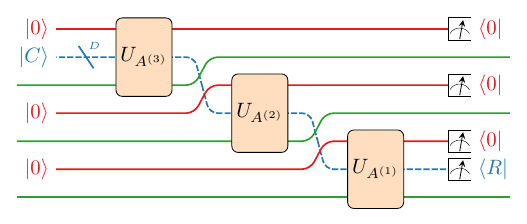}
    \caption{Quantum circuit implementing the MPO representation of a generic quantum Hamiltonian, illustrated for a system with $L=3$ sites.}
    \label{fig:circuit_hamiltonian}
\end{figure}
\begin{figure*}
\centering
\includegraphics[width=\textwidth]{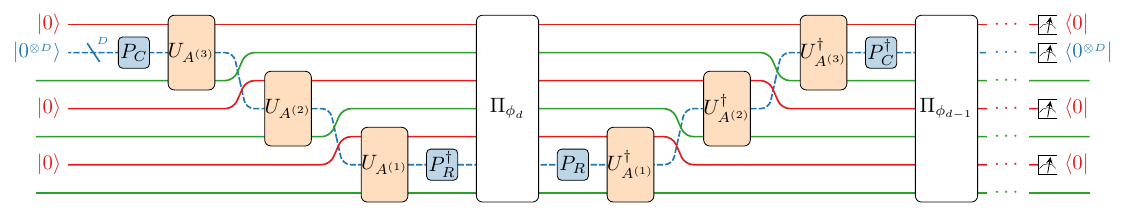}
\caption{MPO block encoding in the quantum eigenvalue transformation circuit without specifying the signal processing gates.}
\label{fig:qubitization_unspecified_processing}
\end{figure*}

Now that we have achieved a unitary block encoding for each $A^{(\ell)}$, we can represent the entire MPO tensor network by a quantum circuit.
Fig.~\ref{fig:circuit_hamiltonian} shows the full block encoding circuit for a generic Hamiltonian $\mathcal{H}$ acting on a system of size $L = 3$ as an example. Note that:
\begin{itemize}
    \item The block encoding circuit needs in total $L$ physical qubits, marked in green, $D$ virtual bond ancillaries, blue dashed, and $L$ dilation ancillaries, marked in red.
    The total number of ancillaries is then $L+D$, and the color scheme matches the one for the equivalent wires of the tensor networks in Fig~\ref{fig:MPO}.
    \item The encoded Hamiltonian $\mathcal{H}$ is rescaled by a factor:
    \begin{equation}\label{eq_def_N_MPO}
        N_{\text{MPO}} = \prod_{\ell=1}^L N_\ell ,
    \end{equation}
    Note that the normalization factors $N_{\ell}$ are typically bigger than $1$, which means that $N_{\text{MPO}}$ grows exponentially with the system size.
    \item To retrieve $\mathcal{H}$, the dilation ancillaries must be all initialized and measured in the state $\ket{0}$.
    The virtual bond ancillaries must also be prepared in state $\ket{C}$ and measured in state $\ket{R}$, representing the column and row vectors $C$ and $R$ previously defined. 
    As a result, a post-selection operation is needed to fully implement $\mathcal{H}$ on a quantum circuit but, remarkably, the probability of measuring the right states is independent of the number of ancillaries \cite{termanova2024tensor}:
    \begin{align} \label{eq_normalization_MPO}
        p_{\text{success}} &= \left\Vert \bra{0}\otimes\bra{R} U \ket{C}\otimes\ket{0}\otimes\ket{\psi_{\text{in}}} \right\Vert^2 \nonumber\\
        &= \left\Vert \frac{\mathcal{H}}{N_{\text{MPO}}} \ket{\psi_{\text{in}}} \right\Vert^2 \nonumber\\
        &= \sum_k \left\vert \frac{\lambda_k}{N_{\text{MPO}}} \right\vert^2 \cdot \left\vert \braket{\lambda_k|\psi_{\text{in}}} \right\vert^2
    \end{align}
    where $\ket{\psi_{\text{in}}}$ is the input state on the physical wires, we have used the eigendecomposition of $\mathcal{H}$: $\mathcal{H} = \sum_k \lambda_k \ket{\lambda_k}\bra{\lambda_k}$ and the rescaling factor was defined in Eq.~\eqref{eq_def_N_MPO}.
    Interestingly enough, the probability is maximal when $\ket{\psi_{\text{in}}}$ has a big overlap with the eigenvector corresponding to the maximal eigenvalue.
    However, the rescaling factor $N_{\text{MPO}}$ affects the success probability with an exponential damping in $L$. This aspect will be further discussed in the following sections.
    
    \item The cost for decomposing each $U_{A^{(\ell)}}$ into one- and two-qubit gates drastically varies for different applications and different native gate sets of the specific hardware. Nonetheless, in the most general case, it grows exponentially with the number of wires involved in each gate, equal to $D+2$, so that the number of two-qubit gates scales as $\mathcal{O}(L\cdot 4^{D})$ \cite{Krol_2022, Rakyta_2022}.
    Returning to the virtual bond dimension formalism and recalling Eq.~\eqref{eq_virtual_bond_dimension}, we can conclude that the gate decomposition cost for the MPO circuit scales linearly with the system size and quadratically with the virtual bond dimension as $\mathcal{O}(L\cdot \chi^2)$.
\end{itemize}

\section{Quantum eigenvalue transformation with MPO block encoding}\label{sec_QET}
We now illustrate the quantum eigenvalue transformation (QET) algorithm \cite{Martyn_2021} as a possible application of our block encoding method.

Given a block encoding $U$ of a Hamiltonian $\mathcal{H}$ and a  polynomial of degree $d$, there exists a sequence of angles $\{\phi_k\}_{k=1,\dots,d}$ so that (in the case of $d$ even):
\begin{equation}\label{eq_eigenvalue_transform_even}
    \prod_{k=1}^{d/2} \left(\Pi_{\phi_{2k-1}} U^{\dagger} \Pi_{\phi_{2k}} U \right)= \begin{pmatrix}
            \text{Poly}_d(\mathcal{H}) & * \\
            * & *
        \end{pmatrix}.
\end{equation}
The signal processing operators $\Pi_{\phi_k}$ are defined as:
\begin{equation}\label{eq_def_signal_processing}
    \Pi_{\phi_k} = e^{-i\phi_k\left(2\ket{\Pi}\bra{\Pi}-I\right)}
\end{equation}
where $\ket{\Pi}\bra{\Pi}$ is the projector onto $\mathcal{H}$: $\bra{\Pi}U\ket{\Pi} = \mathcal{H}$.
Similarly, for odd $d$, Eq.~\eqref{eq_eigenvalue_transform_even} becomes:
\begin{equation}\label{eq_eigenvalue_transform_odd}\hspace*{-0.25 cm}
    \Pi_{\phi_1}U \prod_{k=1}^{(d-1)/2} \left(\Pi_{\phi_{2k}} U^{\dagger} \Pi_{\phi_{2k+1}} U \right)= \begin{pmatrix}
            \text{Poly}_d(\mathcal{H}) & * \\
            * & *
        \end{pmatrix}.
\end{equation}

In the previous sections, we have illustrated how to create a unitary encoding for an MPO Hamiltonian acting on $L$ sites through $L+D$ ancillary qubits. 
To retrieve the Hamiltonian from $U_{A^{(\ell)}}$ depicted in Fig.~\ref{fig:circuit_hamiltonian}, we need to project onto $\ket{0}$ for all the red wires (both initialization and measurement), while for the virtual bond ancillaries we need to initialize the state $\ket{C}$ and measure $\ket{R}$. For the sake of simplicity, we make use of the two state-preparation gates $P_C$ and $P_R$:
\begin{align}
    P_C\ket{0\cdots0} &= \ket{C} \\
    P_R\ket{0\cdots0} &= \ket{R}
\end{align}
to initialize and measure the virtual bond ancillaries in the state $\ket{0\cdots0}$, just like the dilation ancillaries.

In section \ref{sec_circuit_rep}, we have shown how the probability of success in the post-selection process does not depend on the number of ancillaries.
A similar analysis can be done for the full QET circuit:
\begin{align}\label{eq_p_success_qet}
        p_{\text{success}} &= \left\Vert \bra{0}\otimes\bra{R} U_\text{QET} \ket{C}\otimes\ket{0}\otimes\ket{\psi_{\text{in}}} \right\Vert^2 \nonumber\\
        &= \left\Vert \text{Poly}_d\left(\frac{\mathcal{H}}{N_{\text{MPO}}}\right) \ket{\psi_{\text{in}}} \right\Vert^2 \nonumber\\
        &= \sum_k \left\vert\text{Poly}_d\left(\frac{\lambda_k}{N_{\text{MPO}}}\right) \right\vert^2 \cdot \left\vert \braket{\lambda_k|\psi_{\text{in}}} \right\vert^2
\end{align}
Remarkably, this time, the probability of success is maximal when $\ket{\psi_{\text{in}}}$ has a big overlap with the eigenvector $\ket{\lambda_k}$ corresponding to the largest eigenvalue (in magnitude) after the transformation, and not of the original Hamiltonian.
Note also that the rescaling factor $N_{\text{MPO}}$ does not directly suppress the success probability anymore because it now appears in the argument of the polynomial transformation.
Still, the exponential suppression term can squeeze the original eigenvalues towards $0$, so that we would need an exponentially increasing $d$ to distinguish the different $\lambda_k$.

\begin{figure*}
    \centering
\includegraphics{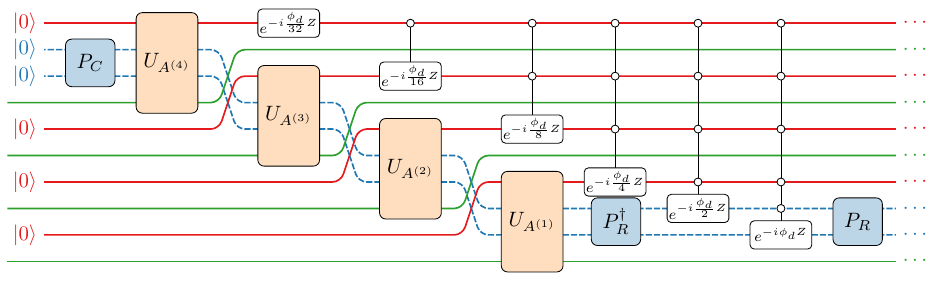}
    \caption{First QET step with $L=4$ and $D=2$. If the Hamiltonian acts on $L$ particles, the first $L-1$ terms of the signal processing circuit can be executed in parallel with the block encoding.}
\label{fig:block_encoding_processing}
\end{figure*}

The overall QET circuit, before specifying how to build the signal processing operator, is visualized in Fig.~\ref{fig:qubitization_unspecified_processing}.

Martyn et al.'s paper \cite{Martyn_2021} contains a first proposal for building the operator $\Pi_{\phi_k}$ with one extra auxiliary qubit. As an example, the mentioned operator would be implemented in the following way for $L=3$ and $D=2$:
\begin{equation} \label{eq_processing_gu}
    \Pi_{\phi_k} = \vcenter{\hbox{ \includegraphics{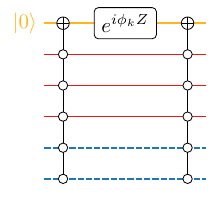}}}
\end{equation}
Note that we always need exactly one extra auxiliary qubit (in yellow), and the signal processing gate acts on all the dilation and virtual bond ancillaries.
Decomposing the two Toffolis requires then $\mathcal{O}\left((L+D)^2\right)$ one- and two-qubit gates, or even $\mathcal{O}\left(L+D\right)$ if we allow a second extra ancillary \cite{Saeedi_2013, Zindorf_2024}.

Here we suggest a possible alternative approach to implement $\Pi_{\phi_k}$ using multi-controlled $Z$-rotations, instead of multi-controlled NOTs, and no extra auxiliary qubit.
For the same example as in Eq.~\eqref{eq_processing_gu}, the circuit reads (see Appendix~\ref{appendix_signal_proc} for a derivation):
\begin{equation}\label{eq_grand_unification_signal_processing}
    \Pi_{\phi_k} = \vcenter{\hbox{\includegraphics{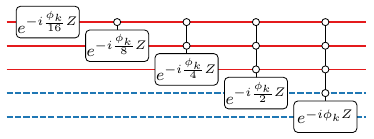}}}
\end{equation}
where the $n$-th gate is controlled by $n-1$ qubits and the corresponding $Z$-rotation has phase $\frac{\phi}{2^{L+D-n}}$.

An $SU(2)$ multi-controlled single-qubit unitary can be decomposed in one- and two-qubit elementary gates with only a linear scaling and no extra ancillaries \cite{Vale_2024}.
By summing all the controlled rotations, we retrieve then a quadratic computational complexity, which matches the first circuit if decomposed without any other ancillary.

Note that the cascading shape of our MPO block encoding allows part of the signal processing circuit to be executed in parallel with the block encoding, with a further reduction of the circuit's depth, as illustrated in Fig.~\ref{fig:block_encoding_processing}.
Given the interchangeability of the ancillaries from the signal processing point of view, we can invert the order of the controlled rotations and start from the bottom wire up, so that also $U^{\dagger}$ can be partially executed in parallel with the following $\Pi_{\phi_k}$.

This decomposition could be practical for devices that have controlled phase rotations instead of cNOTs as native gates. However, it could present a worse scaling if we allow for extra ancillaries.
We are then leaving the choice for the signal processing implementation open since many factors have to be taken into account.

\section{Cost Analysis}\label{sec_cost_analysis}
\begin{figure}[b]
    \centering
    \includegraphics{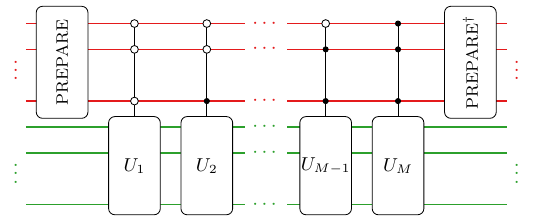}
    \caption{Linear combination of unitaries block encoding technique}
    \label{fig:LCU}
\end{figure}

Before illustrating some possible applications, we would like to discuss the cost of a general QET circuit built with our block encoding algorithm in comparison to the linear combination of unitaries (LCU) technique \cite{Childs_2012, Low_2019, Berry_2015} shown in Fig.~\ref{fig:LCU}.
More specifically, we evaluate the computational cost based on the count of one- and two-qubit gates, and we compare the post-selection success probabilities for the two methods.

The LCU method is based on the insight that every Hamiltonian can be written as a linear combination of unitaries:
\begin{equation}\label{eq_def_LCU}
    \mathcal{H} = \sum_{m=1}^M \alpha_m U_m.
\end{equation}
For concreteness and without loss of generality, we will consider the case where all the $U_M$ are Pauli strings and the $\alpha_m$ are real and positive.
The total number of terms in the summation is denoted as $M$. In the worst case scenario, $M$ scales exponentially with the system size $L$ as $M = \mathcal{O}(4^L)$. 
Still, there are important examples of polynomial or even linear scaling, like the Ising or Heisenberg Hamiltonian. We will first discuss the general case and then return to the specific examples later.

For LCU, one first adds $\lceil\log(M)\rceil$ ancillary qubits to the circuit that need to be prepared in the state \cite{Low_2019}:
\begin{equation}
    \ket{\text{PREP}} = \frac{1}{\sqrt{N_{\text{LCU}}}}\sum_{m=1}^M \sqrt{\alpha_m} \ket{m}
\end{equation}
with the same coefficients $\alpha_m$ as in Eq.~\eqref{eq_def_LCU} and the factor $N_{\text{LCU}}$ defined as:
\begin{equation} \label{eq_def_N_LCU}
    N_{\text{LCU}} = \Vert \alpha \Vert_1 = \sum_{m=1}^M \vert \alpha_m \vert .
\end{equation}
The latter is needed to make sure that $\ket{\text{PREP}}$ is normalized, and this results in a rescaling of the Hamiltonian by a factor $N_{\text{LCU}}$.

This preparation gate, as well as any other multi-qubit gate that we will consider in this section, must be decomposed into one- and two-qubit gates. Generally speaking, this process has a cost that scales exponentially with the number of qubits \cite{Plesch_2011}, which means $\mathcal{O}(M)$.

The next step is known as the selection phase: here, each state $\ket{m}$ controls one of the unitaries $U_m$. We have then $M$ Pauli strings acting on the $L$ system qubits and controlled by the $\lceil\log(M)\rceil$ ancillary qubits.
Supposing that each Pauli gate can be implemented with a constant cost on a quantum computer, we get $\mathcal{O}(L)$ terms for each $U_m$.
Then, for the decomposition of each multi-controlled Pauli, we have two possible choices:
\begin{enumerate}
    \item No extra auxiliaries, which brings a quadratic cost in the number of controls \cite{da_Silva_2022}.
    \item One dirty auxiliary qubit, which allows for a linear scaling with the number of controls \cite{Zindorf_2024} and can be reused by all the multi-controlled Paulis.
    Note that the cited work refers to a $n$-controlled NOT, which can easily be converted into a multi-controlled $Z$ or $Y$ gate with constant cost. 
\end{enumerate}
As a consequence, the cost of the decomposition for the full selection phase scales as $\mathcal{O}(L\cdot M\cdot\log^{\beta}(M))$, where $\beta$ is equal to $1$ or $2$ based on the presence of the extra dirty ancilla.

As a last step, we need to apply the inverse preparation gate on the ancillaries, with a final cost of $\mathcal{O}(2M+L\cdot M\cdot\log^{\beta}(M)) = \mathcal{O}(L\cdot M\cdot\log^{\beta}(M))$.

Considering now some special cases:
\begin{itemize}
    \item In the worst-case scenario, with $M$ scaling as $4^L$, we get $\mathcal{O}(L)$ ancillary qubits and a total cost of $\mathcal{O}(L^{1+\beta}\cdot 4^{L})$ gates. 
    \item For a general linear scaling, we get a total cost for the decomposition of $\mathcal{O}(L^2\cdot\log^{\beta}(L))$ with only $\mathcal{O}(\log(L))$ ancillaries.
    \item The aforementioned Ising and Heisenberg Hamiltonians also present a linear scaling since they are defined as a linear sum of Pauli strings.
    However, because of the nearest-neighbor constraint, they can get even better complexity of $\mathcal{O}(L\cdot\log^{\beta}(L))$, as the Pauli strings $U_m$ are only composed by one or two Pauli gates.
\end{itemize}

Regarding our MPO block encoding, we will only consider its worst-case scenario.
The resulting circuit shown in Fig.~\ref{fig:circuit_hamiltonian} is implemented through $L$ unitaries acting each on $D+2$ wires: decomposing them requires a cost of $\mathcal{O}(L\cdot4^{D})=\mathcal{O}(L\cdot\chi^2)$ at most.
The preparation gates $P_C$ and $P_R$ do not change the formal complexity, as they scale at most as $\mathcal{O}(\chi)$ \cite{Plesch_2011}.
In conclusion, we can state that we always get a linear dependence on the system size $L$ and a quadratic scaling with the virtual bond dimension $\chi$.
It is also worth recalling that the number of ancillary qubits is precisely equal to $L+D$.

Finally, we consider the post-selection cost for both methods.
In section~\ref{sec_circuit_rep}, and more specifically in Eq.~\eqref{eq_normalization_MPO}, we have derived the post-processing success probability for the MPO block encoding.
The same calculations bring us to a similar result for LCU:
\begin{equation} \label{eq_normalization_LCU}
    p_{\text{success}} = \sum_k \left\vert \frac{\lambda_k}{N_{\text{LCU}}} \right\vert^2 \cdot \vert \braket{\lambda_k|\psi_{\text{in}}} \vert^2
\end{equation}
Here, the damping factor $N_{\text{LCU}}$ depends on the sum of the terms $\alpha_m$, as shown in Eq.~\eqref{eq_def_N_LCU}. We can then expect a linear scaling in $M$, compared to the exponential dependence on $L$ of the MPO case.
Still, comparing these methods is not trivial, as the scaling of $M$ with respect to $L$ depends on the system considered. In appendix~\ref{appendix_p_success}, we provide numerical comparisons for the two methods.
It's also worth remembering that the success probability formulas change if we consider the entire QET circuit, as shown in Eq.~\eqref{eq_p_success_qet}. This, of course, stays true also in the LCU case.

Table~\ref{tab:costs} summarizes the MPO and LCU block encoding decomposition costs, the latter with specific examples of $M$ scaling exponentially or linearly with $L$. 
We also added references to the post-selection of the Hamiltonians and the signal processing decomposition cost.
The latter scales only linearly with the number of ancillary qubits if we consider Martyn et al.'s implementation \cite{Martyn_2021} and one extra auxiliary qubit \cite{Zindorf_2024}.

\begin{table*}
\footnotesize
\setlength{\tabcolsep}{5pt} 
\renewcommand{\arraystretch}{1.2} 
\centering
\begin{tabular}{|c |c c c|c|}
    \cline{2-5}
    \multicolumn{1}{c|}{} & \makecell{Block encoding \\ ancillary qubits} & \makecell{Block encoding \\ computational cost} & \makecell{$p_{\text{success}}$ of \\ retrieving $\mathcal{H}$} & \makecell{Martyn et al. \cite{Martyn_2021} signal processing \\ computational cost (1 extra auxiliary)}\\ 
    \hline
    \makecell{MPO Block Encoding} & $L+D$ & $\mathcal{O}\left(L\cdot \chi^2\right)$ & Eq.~\eqref{eq_normalization_MPO} & $\mathcal{O}(L+\log(\chi))$\\[0.5ex]
    \hline
    \makecell{LCU Block Encoding} & $\lceil\log(M)\rceil$ & 
    $\mathcal{O}\left(L\cdot M\cdot\log^{\beta}(M)\right)$  & Eq.~\eqref{eq_normalization_LCU} & $\mathcal{O}(\log(M))$ \\[0.5ex]
    \hline
    \makecell{LCU Block Encoding\\$M=\mathcal{O}(L)$} & $\mathcal{O}\left(\log(L)\right)$ &
    $\mathcal{O}\left(L^2\cdot\log^{\beta}(L)\right)$ & Eq.~\eqref{eq_normalization_LCU} & $\mathcal{O}(\log(L))$\\[0.5ex]
    \hline
    \makecell{LCU Block Encoding\\$M\simeq4^L$} & $2L$ & 
    $\mathcal{O}\left( L^{1+\beta}\cdot  4^L \right)$ & Eq.~\eqref{eq_normalization_LCU} & $\mathcal{O}(L)$ \\[0.5ex]
    \hline
\end{tabular}
\caption{Comparison of the decomposition cost between the LCU and MPO block encoding and signal processing in terms of number of single- and two-qubit gates, of the encoding ancillaries' number and of the post-selection's success probability (without QET). The analysis assumes a cost for decomposing a general unitary on $n$ qubits of $\mathcal{O}(4^n)$ \cite{Krol_2022, Rakyta_2022}.
For the LCU's select phase, we can decompose multi-controlled Pauli gates with a quadratic cost without the use of extra auxiliary qubits \cite{Saeedi_2013,da_Silva_2022} and $\beta=2$, or with one dirty ancilla \cite{Zindorf_2024} and $\beta=1$. We consider Martyn et al.'s  signal processing implementation \cite{Martyn_2021} and the linear $n$-Toffoli decomposition making use of 1 extra auxiliary qubit \cite{Zindorf_2024}. The number of bond ancillary qubits $D$ scales logarithmically with respect to the MPO virtual bond dimension $\chi$ as $D=\lceil \log(\chi) \rceil$.}
\label{tab:costs}
\end{table*}

In summary, given the linear scaling with $L$ and the quadratic dependence on $\chi$, our block encoding turns out to be competitive compared to LCU when considering the circuit size.
If we consider, for example, the aforementioned Ising and Heisenberg models, the corresponding MPO block encoding scales only linearly with $L$ since $\chi$ is constant, as it will be further discussed in section \ref{sec_applications} with other applications.

On the other hand, post-processing can present a lower success probability in the asymptotic limit.
The relatively large number of ancillary qubits can also make the signal processing phase more expensive, but considering the entire QET circuit, we can conclude that this doesn't constitute a limitation since the block encoding still prevails in the overall computational cost.

Finally, we would like to remark that the two techniques are not directly comparable because $M$ and $\chi$ have different dependencies on the system size based on the application.
We will show some example applications in section \ref{sec_applications}.

It is also worth mentioning that this analysis is based on the current technologies not allowing three-qubit gates (or more), and future hardware developments, as well as better multi-controlled unitaries' decomposition techniques, might revise this analysis.

\section{Applications}\label{sec_applications}

This section demonstrates how our block encoding technique can be applied to specific Hamiltonians.
As a first example, we consider the Ising and Heisenberg Hamiltonians, which have a relatively simple MPO representation and a slightly modified version of the XY model with exponentially decaying potential, where we notice a computational speed-up compared to the use of LCU.
Then, we briefly comment on the case of a spinless one-dimensional Fermi-Hubbard Hamiltonian as a simple example of a fermionic system.
Finally, we consider the case where a sum of Pauli strings can be rearranged in a tensor product of sums of local Pauli operators. We also perform an eigenstate filtering procedure for this system through a complete QET circuit.

In appendix~\ref{appendix_p_success}, we report a numerical analysis regarding the post-selection success probabilities for the physical systems considered in this section, comparing both MPO and LCU block encodings.

It's worth mentioning that even if each of these examples shows a constant virtual bond dimension $\chi$, this is not the most general case.
As a matter of fact, $\chi$'s scaling w.r.t the system size $L$ can vary based on the problem considered.
Still, the analysis in section~\ref{sec_cost_analysis} can be applied in every circumstance.

\subsection{Ising model with transverse field}\label{subsec_Ising}

The Ising Hamiltonian with transverse field is defined as follows:
\begin{equation}\label{eq_Ising_def}
    \mathcal{H}_{\text{Ising}} = J\sum_{\ell=1}^{L-1} Z_{\ell} Z_{\ell+1} + g\sum_{\ell=1}^{L} X_{\ell}
\end{equation}
with $Z_{\ell}$ and $X_{\ell}$ Pauli operators acting on the $\ell$-th site.
A suitable MPO decomposition can be achieved by defining the internal $A^{(\ell)}$ tensors as \cite{Ran_2020, Fr_wis_2010}:
\begin{subequations}
\begin{equation}
A^{(\ell)} = \begin{pmatrix}
    I  &  0    &  0  \\
    \sqrt{\vert J \vert}Z_{\ell}  &  0    &  0  \\
    gX_{\ell} &  \text{sign}(J)\sqrt{\vert J \vert}Z_{\ell} &  I
\end{pmatrix}
\end{equation}
where this MPO form was specifically chosen to ensure that $\left\Vert \mathcal{M}\left[A^{(\ell)}\right]\right\Vert \rightarrow 1$ for vanishing $g$ and $J$.

Then, we can ``close'' the tensor chain with the following initial and final vectors of operators, respectively:
\begin{align}
   \tilde{A}^{(1)} &= \begin{pmatrix}
    g X_1 & \text{sign}(J)\sqrt{\vert J \vert} Z_1 & I
\end{pmatrix}, \\ 
    \tilde{A}^{(L)} &= \begin{pmatrix}
    I \\
    \sqrt{\vert J \vert}Z_L \\
    g X_L
\end{pmatrix}.
\end{align}
\end{subequations}
Every local operator $O_{\ell}$ acts on the $\ell$-th lattice site, and $I$ is the $2^L \times 2^L$ identity matrix.
Such a representation can be easily found through a finite state automata picture of the Hamiltonian, as shown in \cite{Fr_wis_2010, Childs_2012}.

One can also verify that the boundary vectors can be obtained via:
\begin{subequations}
\begin{align}
    \begin{pmatrix}
    g X_1 & \text{sign}(J)\sqrt{\vert J \vert} Z_1 & I
\end{pmatrix} &= \begin{pmatrix}
    0 & 0 & I
\end{pmatrix} \cdot A^{(1)}, \\
\begin{pmatrix}
    I \\
    \sqrt{\vert J \vert}Z_L \\
    g X_L
\end{pmatrix} &= A^{(L)} \cdot \begin{pmatrix}
    I \\
    0 \\
    0
\end{pmatrix}
\end{align}
\end{subequations}
with $A^{(1)}$ and $A^{(L)}$ defined as the other $A^{(\ell)}$.
Eq.~\eqref{eq_mpo_def} for the Ising model then becomes:
\begin{equation}\label{eq_mpo_ising}
    \mathcal{H}_{\text{Ising}} = \begin{pmatrix}
    0 & 0 & I
\end{pmatrix} \left[\; \prod_{\ell=1}^L A^{(\ell)}\; \right] \begin{pmatrix}
    I \\
    0 \\
    0
\end{pmatrix}.
\end{equation}

We immediately notice that since the Pauli operators act on single qubits and have dimension $2\times2$, each matrix has dimension $6\times6$ and needs at least 3 qubits to be represented.
As a consequence, all the matrices and vectors in Eq.~\eqref{eq_mpo_ising} gain a new row and/or column in the following way:
\begin{subequations}
\begin{align}
    R &\longrightarrow \begin{pmatrix}
    0 & 0 & 1 & 0
\end{pmatrix} = \bra{10},\\ 
C &\longrightarrow \begin{pmatrix}
    1 \\
    0 \\
    0 \\
    0
\end{pmatrix} = \ket{00},\\
A^{(\ell)} &\longrightarrow \begin{pmatrix}
        I    &  0    &  0  & 0\\
        \sqrt{\vert J \vert}Z_{\ell}  &  0    &  0  & 0\\
        gX_{\ell} &  \text{sign}(J)\sqrt{\vert J \vert}Z_{\ell} &  I  & 0\\
        0    &  0    &  0  & I
    \end{pmatrix}.
\end{align}
\end{subequations}
The state-preparation gates $P_C$ and $P_R$ are the identity operator and $X\otimes I$, respectively.

Among the $3$ qubits that represent each matrix $A^{({\ell})}$, only one has the role of physical qubit (different for every tensor), while the other $2$ are subsequently contracted. Moreover, the meaning of the column and row vectors $C$ and $R$ is now clearer, as they correspond to the initialization of the virtual bond ancillaries in the state $\ket{00}$ and their projection after measurement in the state $\ket{10}$.

\begin{figure}
    \centering
    \makebox[\textwidth][l]{
    \includegraphics[width=1.05\linewidth]{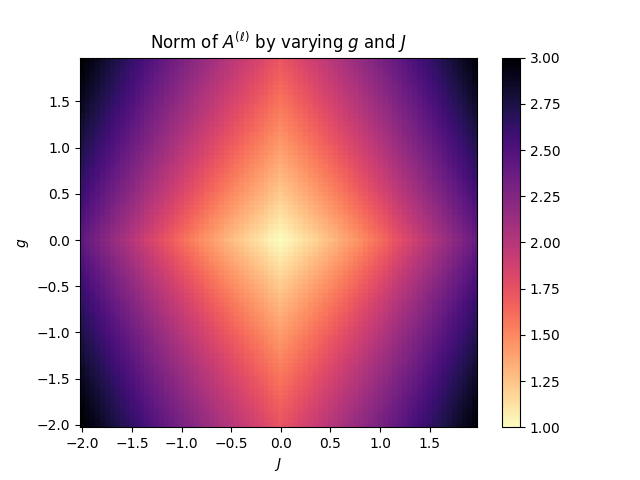}}
    \caption{Normalization factor $N$ (see Eq.~\eqref{eq_normalization_def}) depending on $g$ and $J$.}
    \label{fig:norm_ising}
\end{figure}

Note that to construct a block encoding of $A^{(\ell)}$, we have to normalize it by a factor $N \geq \norm{\mathcal{M}\left[A^{(\ell)}\right]}$, which can be chosen constant for all $\ell$ due to translation invariance. This has the same effect as dividing the Hamiltonian parameters $J$ and $g$ by $N^L$:
\begin{subequations}
\begin{align}
    J &\longrightarrow \frac{J}{N^L} \label{eq_new_J},\\
    g &\longrightarrow \frac{g}{N^L}. \label{eq_new_g}
\end{align}
\end{subequations}
In Fig.~\ref{fig:norm_ising}, we plot the normalization factor $N$ for different values of $g$ and $J$.
We notice that in the limit of vanishing $g$ and $J$, $A^{(\ell)}$ becomes a diagonal matrix with eigenvalues $1$ and $0$, so that we get $\left\Vert \mathcal{M}\left[A^{(\ell)}\right] \right\Vert \rightarrow 1$.

\begin{figure}
    \centering
    \begin{equation*}
        \vcenter{\hbox{ \includegraphics{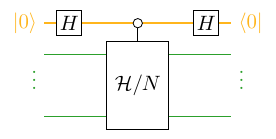}}}
        \hspace*{0.05cm} = \hspace*{0.04cm}
        \vcenter{\hbox{
        \includegraphics{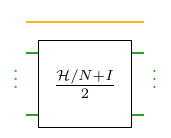}}}
    \end{equation*}
    \caption{One possible strategy for shifting the eigenvalues of $\mathcal{H}$ from the interval $[-1,1]$ to $[0,1]$ that works for every block encoding strategy.}
    \label{fig:positive_Hamiltonian}
\end{figure}

It is worth mentioning that some algorithms, including the eigenstate filtering process that we have applied to the tensor product of Pauli sums, require the eigenvalues of $\mathcal{H}$ not only to have an absolute value smaller than $1$, but also to be positive.
To achieve that, one may add an extra auxiliary qubit and control the entire block encoding process as shown in Fig.~\ref{fig:positive_Hamiltonian} \cite{Martyn_2021}.
Within our specific technique, however, we can add a term $\zeta I$ to Eq.~\eqref{eq_Ising_def} without any extra computational cost by simply changing the MPO definition and initial and final projections:
\begin{subequations}
\begin{align}
    \ket{R} &\longrightarrow \frac{1}{\sqrt{2}}\begin{pmatrix}
    0 & 0 & 1 & 1
\end{pmatrix} = \frac{\bra{10} + \bra{11}}{\sqrt{2}}\\ 
    \ket{C} &\longrightarrow \frac{1}{\sqrt{2}}\begin{pmatrix}
    1 \\
    0 \\
    0 \\
    1
\end{pmatrix} = \frac{\ket{00}+\ket{11}}{\sqrt{2}}\\
A^{(\ell)} &\longrightarrow
\begin{pmatrix}
    I          &  0          &  0  & 0\\
    \sqrt{\vert J \vert}Z_{\ell}   &  0          &  0  & 0\\
    g X_{\ell} &  \text{sign}(J)\sqrt{\vert J \vert} Z_{\ell} &  I  & 0\\
    0          &  0          &  0  & \zeta^{1/L} I
\end{pmatrix}
\end{align}
\end{subequations}

As a consequence, the state preparation operators become:
\begin{subequations}
\begin{align}
    P_R &= \vcenter{\hbox{\includegraphics{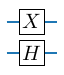}}} \\
    P_C &= \vcenter{\hbox{\includegraphics{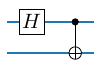}}}
\end{align}
\end{subequations}
Note that, when normalizing the Hamiltonian, we also need to consider an extra factor $2$ given by the normalization of states $\ket{R}$ and $\ket{C}$, so that:
\begin{subequations}
\begin{align}
    J &\longrightarrow \frac{J}{2N^L}, \\
    g &\longrightarrow \frac{g}{2N^L}.
\end{align}
\end{subequations}

In the specific Ising case, we can thus shift the eigenvalues of $\mathcal{H}$ in the interval $[0,1]$ without a significant increase in the computational cost. In other cases, however, adding a new identity operator in the right-bottom corner of $A^{(\ell)}$ could increase the dimension beyond $2^D$, requiring one additional virtual bond ancillary.

\subsection{Heisenberg model}\label{subsec_Heisenberg}
Next, we discuss the Heisenberg model governed by the Hamiltonian:

\begin{align} \label{eq_Heisenberg_def}
    \mathcal{H}_{\text{Heis}} &= \sum_{\ell=1}^{L-1} \left(J_X X_{\ell} X_{\ell+1} + J_Y Y_{\ell} Y_{\ell+1} + J_Z Z_{\ell} Z_{\ell+1}\right) \nonumber\\
    &+ \sum_{\ell=1}^{L}\left( g_X X_{\ell} + g_Y Y_{\ell} + g_Z Z_{\ell}\right).
\end{align}

It can be proven that the corresponding MPO decomposition has a similar structure as for the Ising model with matrices $A^{(\ell)}$ \cite{Yamada_2023, Fr_wis_2010}:
\vspace{3.5cm}
\begin{widetext}
\begin{equation}\label{eq_heisenberg_MPO}
    A^{(\ell)} = \begin{pmatrix}
        I    &  0    &  0  & 0 &  0  \\
        \sqrt{\vert J_X \vert}X_{\ell}  &  0    &  0  & 0 &  0  \\
        \sqrt{\vert J_Y \vert}Y_{\ell}  &  0    &  0  & 0 &  0  \\
        \sqrt{\vert J_Z \vert}Z_{\ell}  &  0    &  0  & 0 &  0  \\
        g_X X_{\ell} + g_Y Y_{\ell} + g_Z Z_{\ell} &  \text{sign}(J_X)\sqrt{\vert J_X \vert} X_{\ell} & \text{sign}(J_Y)\sqrt{\vert J_Y \vert} Y_{\ell} & \text{sign}(J_Z)\sqrt{\vert J_Z \vert} Z_{\ell} &  I  \\
    \end{pmatrix}.
\end{equation}
\end{widetext}

The matrix $A^{({\ell})}$ in Eq.~\eqref{eq_heisenberg_MPO} has dimension $10 \times 10$ and therefore needs to be expanded into a $16 \times 16$ matrix (analogous to the Ising MPO model) to be represented by $4$ qubits.
Moreover, another dilation ancillary is added to build $U_{A^{(\ell)}}$ through a QR decomposition.

As a result, the entire block encoding circuit resembles the Ising example but with $3$ virtual bond ancillaries instead of $2$, given the larger dimensions of $A^{(\ell)}$; the latter needs then to be initialized in the state $\ket{000}$ and measured in the state $\ket{101}$.

\subsection{Exponentially decaying XY model}
The XY model, which restricts the more general Heisenberg model to the XY plane, is particularly interesting. Removing the $2$ rows and columns containing $Z_{\ell}$ results in an MPO matrix of dimensions $8 \times 8$, which can be represented with 3 qubits and no further expansion. Thus, the XY model requires the same computational resources as the Ising model.

Here, more specifically, we will consider a slightly different version of this model, where we will relax the nearest neighbor assumption with an exponentially decaying interaction (and no transverse field, for simplicity):
\begin{equation} \label{eq_XYexp_def}
    \mathcal{H}_{\text{XY}_\gamma} = \sum_{\ell_1=1}^{L} \sum_{\ell_2=\ell_1+1}^{L} e^{-\gamma\vert \ell_1 - \ell_2 \vert} \left(J_X X_{\ell_1} X_{\ell_2} + J_Y Y_{\ell_1} Y_{\ell_2} \right) 
\end{equation}
with $\gamma>0$.

If we wanted to represent such a Hamiltonian with a linear combination of unitaries, we would need to consider $M=\mathcal{O}(L^2)$ terms, which results in a gate decomposition complexity of $\mathcal{O}(L^2\cdot \log^{\beta}(L))$ and $\mathcal{O}(\log(L))$ ancillaries, as discussed in section~\ref{sec_cost_analysis}.

On the other hand, this model has a very simple MPO representation with the same virtual bond dimension as the standard XY model $\chi=4$ \cite{Zaletel_2015}:
\begin{widetext}
\begin{equation}\label{eq_XYexp_MPO}
    A^{(\ell)} = \begin{pmatrix}
        I    &  0    &  0  & 0  \\
        e^{-\gamma}\sqrt{\vert J_X \vert}X_{\ell}  &  e^{-\gamma} I    &  0  & 0   \\
        e^{-\gamma}\sqrt{\vert J_Y \vert}Y_{\ell}  &  0    &  e^{-\gamma} I  & 0  \\
        0 &  \text{sign}(J_Y)\sqrt{\vert J_X \vert} X_{\ell} & \text{sign}(J_Y)\sqrt{\vert J_Y \vert} Y_{\ell} &  I  \\
    \end{pmatrix}.
\end{equation}
\end{widetext}

Our MPO block encoding technique would then be able to represent this Hamiltonian with a number of one- and two-qubit scaling only linearly with the system size $L$. The number of ancillaries would also scale as $\mathcal{O}(L)$ as well as the cost of the signal processing, giving a total cost for a full QET circuit of only $\mathcal{O}(d\cdot L)$, where $d$ is the degree of the polynomial.

It's also worth noticing that with LCU we would need to handle exponentially small coefficients, which might cause practical challenges and require a truncation for very small terms. With the MPO block encoding, on the other hand, the factor $e^{-\gamma}$ is the same in every matrix $A^{(\ell)}$, and we can get an exact representation without any truncation.


\subsection{Spinless Fermi-Hubbard}
We briefly report also the one-dimensional spinless Fermi-Hubbard as another example of Hamiltonian presenting a constant $\chi$. Unlike the other examples shown so far, this is a fermionic Hamiltonian and needs to be embedded in the qubit system by a fermions-to-qubits mapping, e.g., the Jordan-Wigner mapping \cite{Jordan1928berDP}.
The embedded Fermi-Hubbard Hamiltonian looks then like:
\begin{align}
    \mathcal{H}_{\text{FH}} &= J \sum_{\ell=1}^{L-1} \left(S^{+}_{\ell}S^{-}_{\ell+1} + S^{-}_{\ell}S^{+}_{\ell+1}\right) \nonumber \\
    &+ u \sum_{\ell=1}^{L-1} n_{\ell}n_{\ell+1}
\end{align}
where the rising and lowering operators are defined as:
\begin{equation}
    S^{\pm}_{\ell} = \frac{X_{\ell}\mp iY_{\ell}}{2}
\end{equation}
and the number operator is:
\begin{equation}
    n_{\ell} = S^{+}_{\ell}S^{-}_{\ell} .
\end{equation}

The MPO of such a Hamiltonian will then look like:
\begin{widetext}
\begin{equation}
    A^{(\ell)} = \begin{pmatrix}
        I & 0 & 0 & 0 & 0 \\
        \sqrt{\vert J \vert} S^{+}_{\ell} & 0 & 0 & 0 & 0 \\
        \sqrt{\vert J \vert} S^{-}_{\ell} & 0 & 0 & 0 & 0 \\
        \sqrt{\vert u \vert} n_{\ell} & 0 & 0 & 0 & 0 \\
        0 & \text{sign}(J)\sqrt{\vert J \vert} S^{-}_{\ell} & \text{sign}(J)\sqrt{\vert J \vert} S^{+}_{\ell} & \text{sign}(u)\sqrt{\vert u \vert} n_{\ell} & I \\
    \end{pmatrix} .
\end{equation}
\end{widetext}

This model can be further generalized for the spinful case. Still, this simple system suffices to show one possible example of a fermionic Hamiltonian that can be efficiently encoded with our algorithm.

\subsection{Tensor product of sum of Paulis} \label{subsec_Paulis}
Finally, we consider a different Hamiltonian on which we test the entire QET algorithm by applying an eigenstate filtering transformation.
Let us consider the particular case of a tensor product of sums of local Pauli matrices:
\begin{equation}\label{eq_Pauli_def}
\mathcal{H}_{\text{TP}} = \bigotimes_{\ell=1}^L \left(\alpha_{\ell} I + \beta_{\ell} X_{\ell} + \gamma_{\ell}Y_{\ell} + \delta_{\ell}Z_{\ell}\right).
\end{equation}
The MPO representation is straightforward, as each MPO tensor is a one-qubit matrix:
\begin{equation}\label{eq_Pauli_MPO}
    A^{(\ell)} = \left(\alpha_{\ell} I + \beta_{\ell} X_{\ell} + \gamma_{\ell}Y_{\ell} + \delta_{\ell}Z_{\ell}\right) 
\end{equation}
with $D = 0$ virtual bond ancillaries.

Following the eigenstate filtering process from Martyn et al.'s paper \cite{Martyn_2021}, we learn that the Hamiltonian should only have positive eigenvalues.
By following a similar strategy as discussed for the Ising model, we add an element $\zeta I$ to $\mathcal{H}_{\text{TP}}$; the tensors $A^{(\ell)}$ are then modified in the following way:
\begin{equation}
    A^{(\ell)} \longrightarrow \begin{pmatrix}
        \alpha_{\ell} I + \beta_{\ell} X_{\ell} + \gamma_{\ell}Y_{\ell} + \delta_{\ell}Z_{\ell} & 0 \\
        0 & \zeta^{1/L} I
    \end{pmatrix}
\end{equation}
so that each $A^{(\ell)}$ requires now two qubits to be represented, one physical and one constituting the virtual bond ancillary, plus the dilation ancillary. 
The virtual bond ancillary must then be initialized and measured in the state $\ket{C}=\ket{R}=\ket{+}$ so that:
\begin{multline}
    \bigotimes_{\ell=1}^L \left(\alpha_{\ell} I + \beta_{\ell} X_{\ell} + \gamma_{\ell}Y_{\ell} + \delta_{\ell}Z_{\ell}\right) + \zeta I =\\
     \begin{pmatrix}
        I & I  
    \end{pmatrix} \left[ \prod_{\ell=1}^L \begin{pmatrix}
        \alpha_{\ell} I + \beta_{\ell} X_{\ell} \\+ \gamma_{\ell} Y_{\ell} + \delta_{\ell}Z_{\ell} & 0 \\
        0 & \zeta^{1/L} I
    \end{pmatrix} \right] \begin{pmatrix}
        I \\
        I  
    \end{pmatrix},
\end{multline}
which lets us identify both preparation gates $P_C$ and $P_R$ with a Hadamard gate.
Note that the coefficients $\{\alpha_{\ell}, \beta_{\ell}, \gamma_{\ell}, \delta_{\ell}\}$ and $\zeta$ have been redefined in order to absorb the factor 2 derived by the two Hadamard transformations and the normalization factors that make $\norm{\mathcal{M}[A^{(\ell)}]} \leq 1$ for every $\ell$.



We have chosen this example due to its simplicity and to test the correct functionality of the entire QET circuit.

\subsection{Eigenstate filtering}\label{subsubsec_eignstate_filtering}

\begin{figure*}
    \centering
    \subfloat[]{\includegraphics[width=0.5\textwidth]{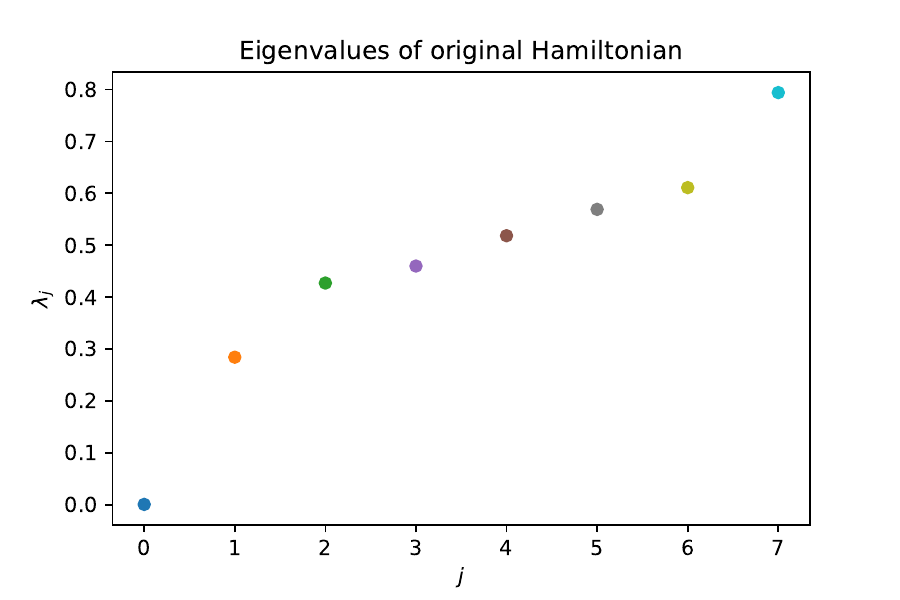}\label{fig:Pauli_original_eigenvalues}}
    \subfloat[]{\includegraphics[width=0.5\textwidth]{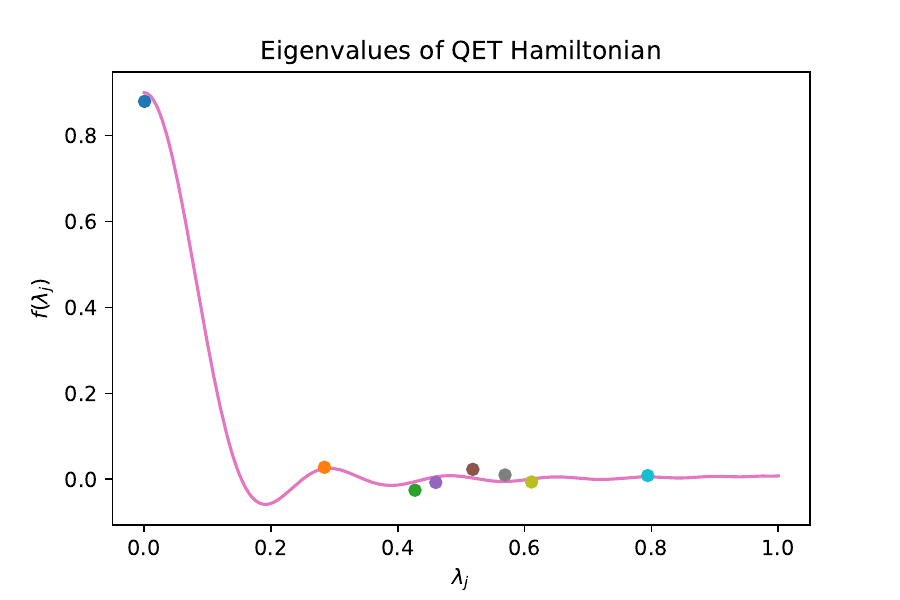}\label{fig:Pauli_qet_eigenvalues}}
    \caption{\ref{fig:Pauli_original_eigenvalues} Original eigenvalues for tensor product of Pauli sums with $L=3$, $\alpha=\{0.7, 1.2, -0.3\}$, $\beta=\{-1, 0.4, 0.5\}$, $\gamma=\{0, 0.3, 0.5\}$, $\delta=\{0.1, 0, 1.2\}$, $\zeta=1.7$ and $N=1.72$. \ref{fig:Pauli_qet_eigenvalues} 
     Filtering function (in pink) plus QET-transformed eigenvalues. Phases were calculated via the \texttt{pyqsp} package with degree $d=30$ and threshold $\Delta_\lambda = 0.1$.}
     \label{fig:eigenstate_filtering}
\end{figure*}
Finally, we test our block encoding algorithm, as well as our signal processing technique, by applying an eigenstate filtering function to the Hamiltonian defined in \eqref{eq_Pauli_def}.
In particular, we want to perform the following transformation over $\mathcal{H}$ \cite{Lin_2020}:
\begin{equation}\label{eq_eigenstate_filtering}
    f_d(x, \Delta_\lambda) = \frac{T_d\left(-1+2\frac{x^2-\Delta_\lambda^2}{1-\Delta_\lambda^2}\right)}{T_d\left(-1+2\frac{-\Delta_\lambda^2}{1-\Delta_\lambda^2}\right)}
\end{equation}
where $\Delta_\lambda$ is the gap between the first and second eigenvalue and $T_d$ is the Chebyshev polynomial of order $d$.
Using the Python package \texttt{pyqsp} \cite{Martyn_2021}, we have been able to get the angles $\phi_k$ better approximating the function $f_k(x, \Delta_\lambda)$ through a QET.
In figure \ref{fig:eigenstate_filtering} one can then see a specific example with the Hamiltonian's initial eigenvalues as well as their final values after the entire QET circuit compared with the target filtering function.

\texttt{pyqsp} was built for working with quantum signal processing and for acting on a specific one-qubit signal operator, which differs slightly from the conventions that we have used. 
Still, it is possible to adapt \texttt{pyqsp}'s angles for our purposes by following the instructions in Appendix A.2 of \cite{Martyn_2021}.

Moreover, a QET process with these phases generates a complex polynomial transformation, whose real part is our target function.
When simulating the circuit, it is therefore necessary to take the hermitian part of the final outcome to get the desired result:
\begin{equation}\label{eq_real_part}
    \text{poly}(\mathcal{H}) = \frac{1}{2}\left(f_{\text{QET}}(\mathcal{H})+f_{\text{QET}}(\mathcal{H})^\dagger\right).
\end{equation}
Such a combination of $f_{\text{QET}}$ and $f_{\text{QET}}^\dagger$ cannot be directly performed on a quantum computer.
For simplicity, we leave the question of how to perform this operation on a quantum computer for future work. To demonstrate the working of the algorithm, we have evaluated it classically.

\section{Conclusion}\label{sec_conclusion}
Our block encoding method is generally applicable since typical Hamiltonians of interest can be represented as matrix product operators.
Moreover, our block encoding technique's gate decomposition cost grows only linearly with the system size and quadratically with the virtual bond dimension, offering a potential advantage compared to LCU.
In section \ref{sec_applications} we have shown one specific example of a circuit size advantage in the case of an exponentially decaying XY model.

The signal processing operator can have different implementations, as we have shown in section~\ref{sec_QET}, and it incurs a cost growing linearly or quadratically with the number of ancillaries, depending on the chosen implementation and the presence of one extra auxiliary qubit \cite{da_Silva_2022, Zindorf_2024, Vale_2024}.

As previously mentioned, each matrix encoded into a larger unitary must have a bounded norm. Encoding directly $\mathcal{H}$ in a larger unitary requires $\norm{\mathcal{H}} \leq 1$.
In our case, we need to normalize each matrix $\mathcal{M}[A^{(\ell)}]$ individually, causing an equivalent rescaling of the Hamiltonian by the factor $N_{\text{MPO}}$ defined by Eq.~\eqref{eq_def_N_MPO}.
The required renormalization is a general issue of block encoding and is not specific to our MPO approach. We remark that the spectral norms of the individual matrices $\mathcal{M}[A^{(\ell)}]$ are typically easier to calculate than $\norm{\mathcal{H}}$. 
However, $N_{\text{MPO}}$ grows exponentially with $L$, which causes an exponentially small success probability for our method in the asymptotic limit.
Nevertheless, in appendix~\ref{appendix_p_success} we have shown how rescaling $\mathcal{H}$ could partially solve this issue when considering finite systems.

Instead, the final success probability for a QET circuit is not directly impacted by the normalization factor, which appears in the polynomial transformation argument, see Eq.~\eqref{eq_p_success_qet}.
Still, an exponential suppression of the eigenvalues might bring some practical disadvantages.
This is currently the biggest limitation of our method, but we have been wondering if a different MPO implementation could help to reduce the norm $\Vert \mathcal{M}[A^{(\ell)}]\Vert$, for example considering gauge symmetries of the MPO.
We have decided to leave this question for future work.

A possible future project might focus on generalizing our protocol to Hamiltonians defined on two-dimensional lattices and their \emph{projected entangled-pair operator} representation, inspired by isometric tensor networks \cite{Zaletel_2020} to construct the unitary dilations. This is particularly promising since the qubit topologies of modern quantum computers are often likewise two-dimensional grids.

Moreover, the use of adaptive circuits might further reduce the size and depth of the MPO block encoding, as inspired by \cite{smith2024constantdepthpreparationmatrixproduct, sahay2024classifyingonedimensionalquantumstates,stephen2024preparingmatrixproductstates}.

Our work should then be considered as an introductory publication presenting a new and promising technique, which will need to be further studied in the future, together with other new and interesting applications.

Finally, we acknowledge the existence of a similar work that was made publicly available during the review process of this paper \cite{termanova2024tensor}, which uses numerical optimizations to find an approximation with reduced effective bond dimensions.

\section*{Acknowledgements}
M. Nibbi acknowledges funding by the Munich Quantum Valley, section K5 Q-DESSI. The research is part of the Munich Quantum Valley, which is supported by the Bavarian state government with funds from the Hightech Agenda Bayern Plus.

\appendix
\section{Signal processing circuit}\label{appendix_signal_proc}
In this section, we prove that the circuits drawn in Eqs.~\eqref{eq_grand_unification_signal_processing} and \eqref{eq_processing_gu} are equivalent.
First of all, we recall the definition of the signal processing operator in Eq.~\eqref{eq_def_signal_processing}. Without loss of generality, we can consider the case in which  $\ket{\Pi}=\ket{0}^{\otimes n}$, where $n$ is the total number of ancillary qubits.

The matrix form of $\Pi_{\phi}$ is:
\begin{equation}\label{ref_matrix}
\Pi_{\phi} = \begin{pmatrix}
        e^{-i\phi} & & & &\\
         & e^{i\phi} & & &\\
         & & e^{i\phi} & &\\
         & & & \ddots & \\
         & & & & e^{i\phi}
    \end{pmatrix}.
\end{equation}
It follows directly that Eq.~\eqref{eq_grand_unification_signal_processing} holds, since the state $\ket{0}^{\otimes n}$ is the only one that gains a phase $e^{-i\phi}$, while the other states acquire the phase $e^{i\phi}$.

The general form of our signal processing circuit, without distinguishing between dilation and virtual bond ancillaries, equals:
\begin{equation}\label{eq_def_processing_new_general}
    \hspace*{-0.58cm}\vcenter{\hbox{ \includegraphics{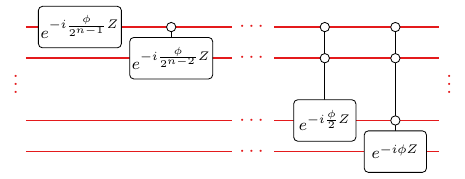}}}
\end{equation}

Let us consider the $q$-th operator in the circuit \eqref{eq_def_processing_new_general}. This operator has $q-1$ controls and a rotation phase of $\frac{-\phi}{2^{n-q}}$.
Its matrix expression, considering the tensor multiplication with $n-q$ identities as well, is:
\begin{equation}\label{eq_q_processing_gate}
    \begin{pmatrix}
        \overbrace{\begin{bmatrix}
            e^{-i\frac{\phi}{2^{n-q}}} &  \\
            & \ddots
        \end{bmatrix}}^{2^{n-q}} & & \\
         & \overbrace{\begin{bmatrix}
            e^{i\frac{\phi}{2^{n-q}}} &  \\
            & \ddots
        \end{bmatrix}}^{2^{n-q}} & \\
         & & \overbrace{\begin{bmatrix}
            1 & \\
            & \ddots \\
        \end{bmatrix}}^{2^{n-q}\left(2^q-2\right)}
    \end{pmatrix}
\end{equation}

By multiplying all the matrices \eqref{eq_q_processing_gate} with $q$ from 1 to $n$, we get then  the following:
\begin{widetext}
\begin{equation}
    \begin{pmatrix}
        \displaystyle \prod_{q=1}^{n} e^{-i\frac{\phi}{2^{n-q}}} & & & & &\\
        & \displaystyle e^{i\phi}\prod_{q=1}^{n-1} e^{-i\frac{\phi}{2^{n-q}}} & & & &\\
        & & \overbrace{\begin{bmatrix}
            \displaystyle e^{i\frac{\phi}{2}}\prod_{q=1}^{n-2} e^{-i\frac{\phi}{2^{n-q}}} & \\[1em]
            & \ddots
            \end{bmatrix}}^{2} & & &\\
        & & & \overbrace{\begin{bmatrix}
            \displaystyle e^{i\frac{\phi}{4}}\prod_{q=1}^{n-3} e^{-i\frac{\phi}{2^{n-q}}} & \\[1em]
            & \ddots
            \end{bmatrix}}^{4} & &\\
        & & & & \ddots &\\
        & & & & & \overbrace{\begin{bmatrix}
            e^{i\frac{\phi}{2^{n-1}}} & \\
            & \ddots
        \end{bmatrix}}^{2^{n-1}} 
    \end{pmatrix}
\end{equation}
\end{widetext}
Apart from the first entry, we can consider $n$ separate diagonal blocks.
Taking $1\leq k \leq n$, the $k$-th block has dimension $2^{k-1}$ and entries equal to $e^{i\frac{\phi}{2^{k-1}}}\prod_{q=1}^{n-k} e^{-i\frac{\phi}{2^{n-q}}}$.
We want to prove that each one of these does not depend on $k$ and, more specifically, is equal to $e^{i\frac{\phi}{2^{n-1}}}$: \begin{align}\label{eq_ref_block}
    e^{i\frac{\phi}{2^{k-1}}}\prod_{q=1}^{n-k} e^{-i\frac{\phi}{2^{n-q}}} &= e^{i\frac{\phi}{2^{k-1}}} e^{-i\phi\sum_{q=1}^{n-k}\frac{2^q}{2^{n}}} \nonumber \\
    &= e^{i\frac{\phi}{2^{k-1}}}e^{-i\phi\frac{2-2^{n-k+1}}{2^n(1-2)}} \nonumber\\[1em]
    &= e^{i\frac{\phi}{2^{k-1}}} e^{-i\phi\left(\frac{1}{2^{k-1}}-\frac{1}{2^{n-1}}\right)} \nonumber\\[1em]
    &= e^{i\frac{\phi}{2^{n-1}}}
\end{align}
\\\\
The first block, on the other hand, has the following entry:
\begin{align}\label{eq_first_block}
    \prod_{q=1}^{n} e^{-i\frac{\phi}{2^{n-q}}} &= e^{-i\phi \sum_{q=1}^{n} \frac{2^q}{2^{n}}} \nonumber\\
    &= e^{-i\phi\frac{2-2^{n+1}}{2^n(1-2)}} \nonumber\\[1em]
    &= e^{i\phi\left(\frac{1}{2^{n-1}}-2\right)}
\end{align}

By multiplying both equations \eqref{eq_ref_block} and \eqref{eq_first_block} by a global phase $e^{i\phi\left(1 + \frac{1}{2^{n-1}}\right)}$ we get then the reference matrix from equation \eqref{ref_matrix}, which proves that the two circuits are equivalent up to a global phase.

\section{Post-selection success probability for systems of interest}\label{appendix_p_success}

\begin{figure*}
    \centering
    \makebox[\textwidth][c]{\includegraphics[width=1.15\textwidth]{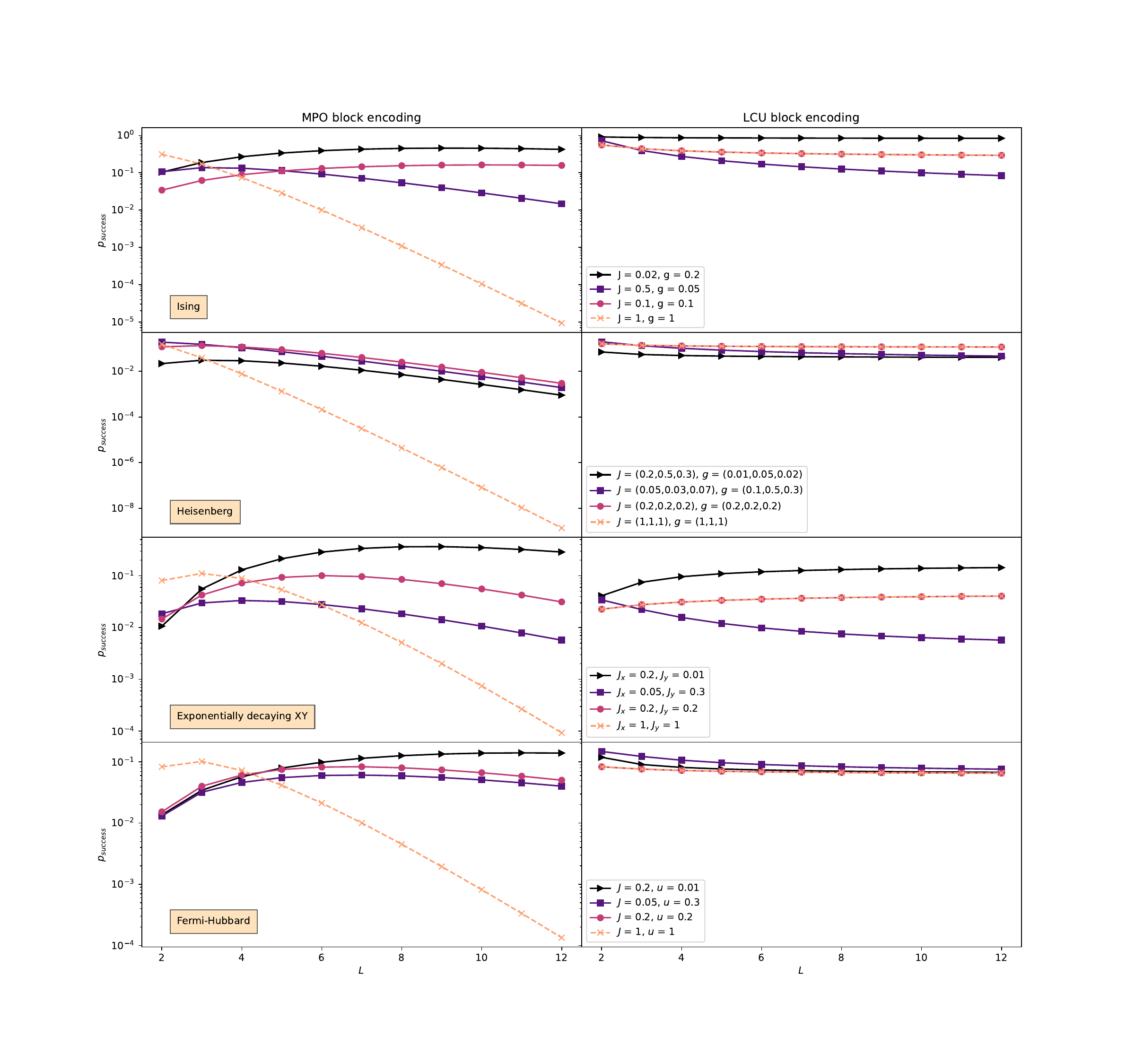}}
     \caption{Success probabilities for LCU and MPO block encodings, given by Eq.~\eqref{eq_normalization_LCU} and Eq.~\eqref{eq_normalization_MPO}. The physical systems are, in order, the Ising, Heisenberg, exponentially decaying XY (with $\gamma=0.3$), and Fermi-Hubbard models.
     The initial state $\ket{\psi_{\text{in}}}$ has been chosen in all cases equal to $\ket{+}^{\otimes n}$. A rescaling of the Hamiltonian parameters can drastically change the probability of success for the MPO block encoding, while it does not have any impact on LCU.}
    \label{fig:p_success_all}
\end{figure*}

In this section, we provide a numerical analysis concerning the probability of success for both MPO and LCU block encodings (without considering the entire QET circuit).
The two methods are summarized respectively in Eq.~\eqref{eq_normalization_MPO} and Eq.~\eqref{eq_normalization_LCU}, where the 
rescaling factors $N_{\text{MPO}}$ and $N_{\text{LCU}}$ are defined by Eq.~\eqref{eq_def_N_MPO} and Eq.~\eqref{eq_def_N_LCU}.
The post-selection success probability scales then exponentially in $L$ for the MPO block encoding and linearly in $M$ for the LCU method.

We remark that the number of unitaries $M$ of the LCU decomposition can scale exponentially in $L$ in the worst-case scenario but also linearly for many systems of interest, such as Ising, Heisenberg, etc.
This means that our method, in the asymptotic limit, presents a much lower success probability compared to LCU in many circumstances.

However, in section~\ref{sec_applications}, and more specifically in Fig.~\ref{fig:norm_ising}, we discussed how rescaling the Hamiltonian might help to reduce the normalization factor so that $N_{\text{MPO}}\rightarrow 1$. 
This trick doesn't work in the LCU case: the same rescaling would also be applied to the eigenvalues of $\mathcal{H}$ so that the success probability would stay constant.

In Fig.~\ref{fig:p_success_all}, we have plotted the post-selection probabilities for the physical systems presented in section~\ref{sec_applications}. We considered different regimes for every Hamiltonian, and we have fixed small parameters to reduce $N_{\text{MPO}}$ as much as possible.
We have also considered one case for every system with larger parameters to show the sharper decay that is encountered when $\mathcal{H}$ is not adequately rescaled.
The equivalent LCU graph doesn't show any difference by an equal rescaling of the Hamiltonian's parameters, which confirms our previous point.

We can argue that the MPO block encoding shows an exponential suppression on the asymptotic limit.
However, we are still able to get functional probabilities perfectly comparable to the LCU method when considering a finite number of lattice sites and by applying a smart rescaling of $\mathcal{H}$.

The exponential scaling of $p_{\text{success}}$ still represents the biggest limitation of our method.
However, we have shown that for a wide range of circumstances, we are still able to find probabilities that are comparable to the LCU block encoding.

\bibliography{bibliography}

\end{document}